\documentclass[12pt,preprint]{aastex}
\usepackage{lscape,graphicx,rotating}

\newcommand{\Swift}{\textit{Swift}}

\newcommand{\Rc}{\textit{R$_{\mathrm{C}}$}}
\newcommand{\Ic}{\textit{I$_{\mathrm{C}}$}}
\newcommand{\Vc}{\textit{V$_{\mathrm{C}}$}}
\newcommand{\zG}{\textit{z}}
\newcommand{\iG}{\textit{i}}
\newcommand{\gG}{\textit{g}}
\newcommand{\ip}{\textit{i$^{\prime}$}}
\newcommand{\zp}{\textit{z$^{\prime}$}}
\newcommand{\gp}{\textit{g$^{\prime}$}}

\begin{document}

\title{Dark Bursts in the \Swift\ Era: The Palomar 60\,inch-\Swift\
Early Optical Afterglow Catalog}

\author{S.~B.~Cenko\altaffilmark{1,2}, J.~Kelemen\altaffilmark{3}, 
  F.~A.~Harrison\altaffilmark{1}, D.~B.~Fox\altaffilmark{4}, 
  S.~R.~Kulkarni\altaffilmark{5}, M.~M.~Kasliwal\altaffilmark{5}, 
  E.~O.~Ofek\altaffilmark{5}, A.~Rau\altaffilmark{5}, 
  A.~Gal-Yam\altaffilmark{6}, D.~A.~Frail\altaffilmark{7}
  and D.-S.~Moon\altaffilmark{8}}
  
\altaffiltext{1}{Space Radiation Laboratory, MS 220-47, California 
  Institute of Technology, Pasadena, CA 91125}
\altaffiltext{2}{Department of Astronomy, 601 Campbell Hall, University of
  California, Berkeley, CA 94720}
\altaffiltext{3}{Konkoly Observatory, H-1525, Box 67, Budapest, Hungary}
\altaffiltext{4}{Department of Astronomy \& Astrophysics, 525 Davey
  Laboratory, Pennsylvania State University, University Park, PA 16802}
\altaffiltext{5}{Department of Astronomy, Mail Stop 105-24, California
  Institute of Technology, Pasadena, CA 91125}
\altaffiltext{6}{Benoziyo Center for Astrophysics, Weizmann
  Institute of Science, Rehovot 76100, Israel}
\altaffiltext{7}{National Radio Astronomy Observatory, P.O.~Box 0, 1003
  Lopezville Road, Socorro, NM 87801}
\altaffiltext{8}{Department of Astronomy and Astrophysics, University of 
  Toronto, 50 St.~George Street, Toronto, ON M5S 3H4, Canada.}

\email{cenko@srl.caltech.edu}

\slugcomment{Submitted to \apj}

\shorttitle{P60 Afterglow Catalog}
\shortauthors{Cenko {\it et al.}}

\begin{abstract}
We present multi-color optical observations of
long-duration $\gamma$-ray bursts (GRBs) made over a three year period with 
the robotic Palomar 60\,inch telescope (P60).  Our sample consists of all 29 
events discovered by \Swift\ for which P60 began observations less than one 
hour after the burst trigger.  We were able to recover $80\%$ of the optical 
afterglows from this prompt sample, and we attribute this high efficiency to 
our red coverage.  Like \citet{mmk+08}, we find that a significant fraction 
($\approx 50\%$) of \Swift\ events show a suppression of the optical flux with 
regards to the X-ray emission (so-called ``dark'' bursts).  Our multi-color
photometry demonstrates this is likely due in large part to extinction in the 
host galaxy.  We argue that previous studies, by selecting only the brightest 
and best-sampled optical afterglows, have significantly underestimated the 
amount of dust present in typical GRB environments.  
\end{abstract}


\section{Introduction}
\label{sec:intro}
The launch of the \Swift\ $\gamma$-Ray Burst (GRB) Explorer \citep{gcg+04} in 
2004 November has ushered in a new era in the study of GRB 
afterglows.  \Swift\ offers a unique combination of event rate ($\sim 
100$\,yr$^{-1}$; almost an order of magnitude increase over previous missions) 
and precise localization ($\sim 3\arcmin$ radius error circles are distributed
seconds after the burst, and refined to $\sim 3\arcsec$ minutes later).  
The on-board X-ray Telescope (XRT; \citealt{bhn+05}) and UV-Optical Telescope 
(UVOT; \citealt{rkm+05}), together with the rapid relay of these 
precise localizations to ground-based observers, has enabled an unprecedented 
glimpse into the time period immediately following the prompt emission over
a broad frequency range.

Observations of X-ray afterglows with the XRT have generated particular
interest in recent years.  In the pre-\Swift\ era, X-ray observations were
limited to hours or days after the prompt emission, and were often poorly
sampled compared with the optical and radio bandpasses.  Routine XRT
observations of \Swift\ GRBs beginning at early times have revealed a central 
engine capable of injecting energy into the forward shock at times well
beyond the duration of the prompt emission (e.g., \citealt{brf+05,zfd+06}).
This discovery has had a profound effect on our understanding of progenitor 
models.

While the X-ray afterglow is currently a well-explored phase space,
comparatively few analogous studies have been performed in the optical
bandpass.  \citet{bkf+05} first suggested that \Swift\ optical afterglows 
were 1.8\,mag fainter in the $R$-band than pre-\Swift\ events (at a common
epoch of 12\,hours after the burst).  Likewise, \citet{rsf+06} found that
only 6 of the first 19 \Swift\ bursts with prompt ($\Delta t \lesssim 100$\,s)
UVOT coverage yielded optical afterglow detections.  Since then, explaining the
faintness of \Swift\ optical afterglows has remained one of the 
outstanding questions in the field.

One clear contributor is distance: the median redshift of \Swift\ events 
($\langle z_{Swift} \rangle \approx 2.0$)\footnote{Calculated from J.~Greiner's
compilation at http://www.mpe.mpg.de/\~{}jcg/grbgen.html.} is significantly 
larger than the pre-\Swift\ sample ($\langle z_{\mathrm{pre-}Swift} \rangle = 
1.1$; \citealt{bkf+05,jlf+06}).  In a comprehensive literature-based study of 
the brightest, best-studied \Swift\ afterglows, \citet{kkz+07} find properties 
broadly similar to pre-\Swift\ events, after applying a cosmological 
k-correction.  

On the other hand, \citet{mmk+08} have recently presented a sample of 63
GRBs observed in the optical ($r^{\prime}$-band) with the robotic 2\,m 
Liverpool Telescope and Faulkes Telescopes (North and South).  The selection 
criteria for including a burst in their sample is never explicitly stated, and 
several non-\Swift\ bursts are included, making a direct comparison with the 
results of \citet{kkz+07} difficult.  However, \citet{mmk+08} do not exclude the
significant fraction of events without optical detections from their analysis, 
providing a more unbiased look at optical afterglow properties.  
By measuring the ratio of optical to X-ray flux at a common time, these 
authors find that roughly half of the GRBs in their sample exhibit a relative 
suppression of the optical flux inconsistent with our standard picture of 
afterglow emission (e.g., \citealt{spn98}), so-called ``dark'' bursts 
\citep{jhf+04}.  This finding suggests that distance alone cannot explain the 
faintness of \Swift\ optical afterglows.  

Several other possibilities have been suggested to explain optically dark
GRB afterglows.  Undoubtedly some GRBs, like GRB\,050904 \citep{hnr+06,kka+06},
originate from such large redshifts ($z \gtrsim 6$) that Ly-$\alpha$ absorption
in the inter-galactic medium (IGM) completely suppresses the optical flux
\citep{lr00}.  Alternatively, late-time energy injection from the central
engine, manifested as bright X-ray flares and/or extended periods of shallow
decay, may be artificially increasing the X-ray flux, leading to spurious claims
of optically dark GRBs \citep{mmk+08}.

One final possibility is extinction native to the GRB host galaxy.  As a 
population, long-duration GRB host galaxies exhibit extremely large neutral H 
column densities (e.g., \citealt{hmg+03,bpc+06}), typically falling at $\log 
N_{H} > 20.3$\,cm$^{-2}$ (so-called Damped Ly-$\alpha$, or DLA systems; 
\citealt{wgp05}).  And within their hosts, GRBs trace the blue light from hot 
young stars in the disk even more closely than core-collapse supernovae
\citep{bkd02,fls+06}.  Both findings are consistent with the observed 
association between long-duration GRBs and massive star death (e.g., 
\citealt{wb06}).

In spite of these expectations, relatively few GRB afterglows to date exhibit 
signs of large host galaxy extinction (e.g., \citealt{cbm+07,rvw+07,tlr+08}).  
\citet{kkz+07} find only a modest amount of dust ($\langle A_{V} \rangle = 
0.20$\,mag) for the 15 events in their ``golden'' sample, an identical value 
found from an analogous study of pre-\Swift\ afterglows \citep{kkz06}.
The primary drawback of such studies, however, is the large and uncertain role
of selection effects: by including only the brightest, best-sampled optical 
afterglows, \citet{kkz+07} may be preferentially selecting those events in
low-extinction environments.  Understanding these selection effects is one
of the primary goals of this work.

The Palomar 60\,inch telescope (P60) is a robotic, queue-scheduled facility
dedicated to rapid-response observations of GRBs and other transient
events \citep{cfm+06}.  With a response time of $\Delta t \lesssim 3$\,min
and a limiting magnitude of $R \gtrsim 20.5$ (60\,s exposure), the P60 
aperture is well suited to detect most \Swift\ optical afterglows \citep{as07}. 
In addition, with a broadband filter wheel providing coverage from the near-UV 
to the near-IR, P60 can also provide multi-color data on the afterglow 
evolution.

In this work, we present the P60-\Swift\ Early Optical Afterglow sample: 
29 unambiguously long-duration GRBs detected by the \Swift\
Burst Alert Telescope (BAT; \citealt{bbc+05}) with P60 observations beginning
at most one hour after the burst trigger time.  This sample offers two 
distinct advantages over previous efforts to understand the optical
afterglow emission from GRBs.  First and foremost, our study enforces a
strict selection criterion independent of the optical afterglow properties, 
and therefore will allow us to study the properties of the \Swift\ population 
in a relatively unbiased manner.  Secondly, nearly all events contain 
multi-color ($\gp\,\Rc\,\ip\,\zp$) observations that allow us to evaluate the 
importance of host galaxy extinction for a fraction of our sample.  Altogether,
we aim to discriminate between the competing hypotheses proffered to explain
dark GRB afterglows in the \Swift\ era.

Throughout this work, we adopt a standard $\Lambda$CDM cosmology with 
$h_{0}$ = 0.71\,km s$^{-1}$ Mpc$^{-1}$, $\Omega_{\mathrm{m}} = 0.27$, and 
$\Omega_{\Lambda} = 1 - \Omega_{\mathrm{m}} = 0.73$ \citep{sbd+07}.  We define the 
flux density power-law temporal and spectral decay indices $\alpha$ and 
$\beta$ as $f_{\nu} \propto t^{-\alpha} \nu^{-\beta}$ (e.g., \citealt{spn98}).  
All errors quoted are 1 $\sigma$ (i.e., $68\%$) confidence intervals unless 
otherwise noted.  

\section{Observations and Data Reduction}
\label{sec:p60catalog}
The P60-\Swift\ Early Optical Afterglow Catalog is shown in 
Table~\ref{tab:p60catalog}.  We have included here all optical afterglows of 
events localized by \Swift\ in the three year period from 2005 April 1 -- 
2008 March 31 (roughly coinciding with the beginning of real-time GRB alerts 
and narrow-field instrument follow-up) for which we began P60 observations 
within one hour after the BAT trigger.  

All P60 data were reduced in the IRAF\footnote{IRAF is distributed by the
National Optical Astronomy Observatory, which is operated by the
Association for Research in Astronomy, Inc., under cooperative agreement with
the National Science Foundation.} environment using our custom real-time
reduction pipeline \citep{cfm+06}.  Where necessary, co-addition was performed
using SWarp\footnote{See http://terapix.iap.fr/soft/swarp.}.  For the vast 
majority of events, magnitudes were calculated using aperture photometry with
the inclusion radius roughly matched to the stellar PSF FWHM.  For the few
events with either extremely crowded fields or variable, elevated backgrounds
(due to nearby bright stars or the moon), image subtraction was performed 
using the ISIS package \citep{al98}.

Photometric calibration was performed relative to the SDSS data release 6
\citep{aaa+08} where possible, typically resulting in rms variations of 
$\lesssim 0.05$\,mag in all filters.  For those fields without Sloan coverage, 
we made use of the calibration files provided by A.~Henden\footnote{Available 
via ftp at ftp.aavso.org.} when available, resulting in similar quality 
calibrations to the SDSS.  The remaining events were calibrated relative to the
USNO-B1 catalog\footnote{See http://www.nofs.navy.mil/data/fchpix.}, resulting 
in significantly poorer zero point fits.  Particularly in the \gG, \zG, and 
\zp\ filters, the rms errors for these events could be quite large 
($\sim 0.6$\,mag).  Photometric and instrumental errors have been added in 
quadrature to obtain the results presented in Table~\ref{tab:p60catalog}.

Filter transformations (either from the Johnson-Kron-Cousins Vega system to the
SDSS AB system, or vice versa) were made using the results from \citet{jga06}.
Throughout this work, the Gunn \gG\ and \zG\ filters have been calibrated
relative to the SDSS \gp\ and \zp\ filters, and their corresponding magnitudes
are reported in the AB system.  The Gunn \iG\ filter, used for some early
observations in 2005, was found to best match the Cousins \Ic\ filter, and hence
is reported on the Vega system.  The remaining filters have magnitudes
reported in their native photometric system (i.e., Vega for \Vc, \Rc, and \Ic,
and AB for \ip\ and \zp).  A summary of the relevant photometric calibration 
and appropriate zero point for flux conversion can be found in 
Table~\ref{tab:p60filters}.  Full throughput curves for all P60 filters can 
be found in \citet{cfm+06}.

Finally, we note that the magnitudes reported in Table~\ref{tab:p60catalog}
have not been corrected for Galactic extinction along the line-of-sight.  
For all subsequent figures and analysis, this correction has been applied 
using the dust extinction maps of \citet{sfd98} and the Milky Way extinction 
curve from \citet{ccm89}.  For most bursts, the extinction correction was 
quite small [$\langle E(B-V) \rangle = 0.04$\,mag], although a few events
were subjected to large column densities [e.g., GRB\,060110: $E(B-V) = 
0.97$\,mag]

\section{Analysis}
\label{sec:early}
The standard theoretical paradigm to explain GRBs is the relativistic
fireball model (e.g., \citealt{p05}).
In the case of long-duration GRBs, accretion onto the black hole remnant of
massive star core-collapse powers an ultra-relativistic outflow of matter
and/or radiation \citep{w93}.  Shocks and/or instabilities within the 
outflow generate the prompt $\gamma$-rays (i.e., internal shocks).  The 
afterglow emission, on the other hand, is powered by electrons in the
circumburst medium accelerated by the outgoing blast wave (i.e., external
shocks).  The resulting synchrotron spectrum and light curve are well described
by a series of broken power-laws \citep{gs02},
with the break frequencies determined not only by properties of the outflow
($E$, $\theta$, etc.), but also by the nature of the circumburst medium.  In 
what follows we attempt to understand the early optical afterglow phase in the 
context of this model.

\begin{figure}[t!]
  \centerline{\plotone{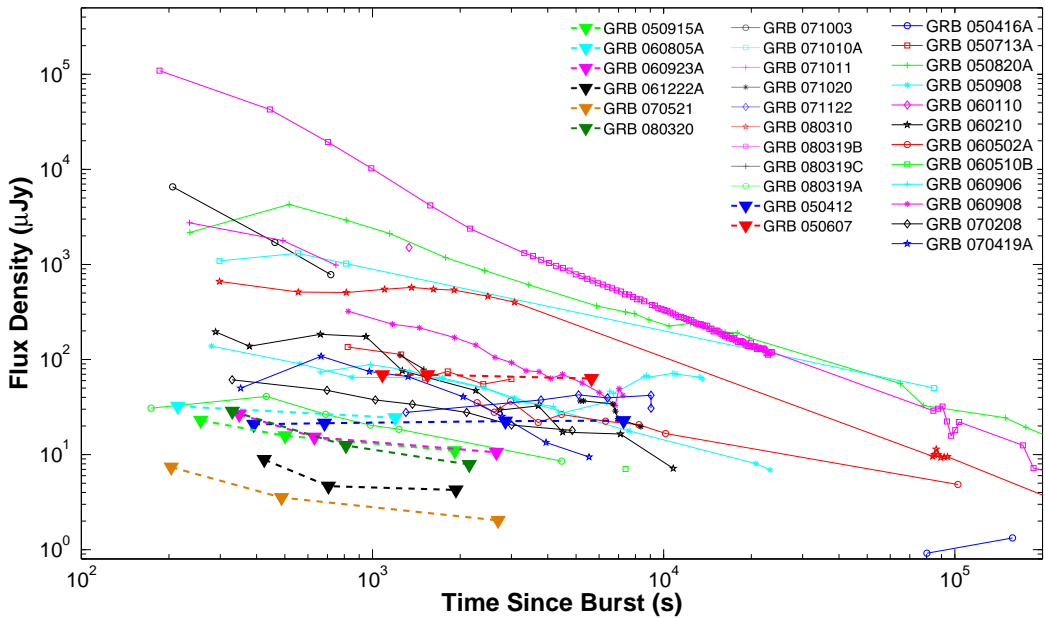}}  
  \caption[The P60-\Swift\ early optical afterglow sample]
  {The P60-\Swift\ early optical afterglow sample.  We plot here \Rc-band
  light curves or upper limits for all 29 events in the P60-\Swift\
  Early Afterglow Sample.  With the exception of GRB\,050607, the upper limits
  fall securely at the very faint end of the distribution (see also
  Fig.~\ref{fig:agdistro}).}
\label{fig:early:all}
\end{figure}

The \Rc-band optical light curves (and upper limits) for all 29 events in the
P60-\Swift\ Early Optical Afterglow Sample are shown in 
Figure~\ref{fig:early:all}.  For all events with P60 optical 
detections, we have simultaneously fit both the spectral and temporal evolution 
of the light curve, assuming a power-law spectrum and either a single or broken 
power-law temporal evolution.  The results of this analysis are shown in
Table~\ref{tab:early}. 

Because afterglow emission is a broadband phenomenon, multi-wavelength
observations can often provide important constraints that would be overlooked
by considering only a single bandpass.  We have therefore obtained XRT
light curves from the on-line \Swift-XRT light curve 
repository\footnote{See http://www.swift.ac.uk/xrt\_curves.} \citep{ebp+07}.
We converted the 0.3--10\,keV fluxes to flux densities at a nominal 
energy of 2\,keV assuming a power-law X-ray spectrum with indices provided
in the Gamma-Ray Burst Coordinate Network (GCN)\footnote{See 
http://gcn.gsfc.nasa.gov/gcn3\_archive.html.} circulars.  We then fit the
temporal decay of each X-ray light curve, assuming either a single or
broken-power law model.  

With these results in hand, we now move on to explore the anomalously large P60
detection efficiency (\S~\ref{sec:deteff:obs}); the relationship between
X-ray and optical flares (\S~\ref{sec:xrayflares}); the brightness
and luminosity distribution of \Swift\ optical afterglows
(\S~\ref{sec:luminosity:obs}); and optically dark bursts in the \Swift\ era
(\S~\ref{sec:dark:obs}).


\subsection{Detection Efficiency}
\label{sec:deteff:obs}
The most striking feature in Table~\ref{tab:early} is the large fraction 
of P60 detected afterglows: of the 29 events in the sample, P60 detected
22 (76$\%$).  This stands in stark contrast with the 32$\%$ 
afterglow detection efficiency of the UVOT \citep{rsf+06} and even 
exceeds the 50$\%$ value reported by the larger Liverpool and Faulkes 
telescopes \citep{mmk+08}.  For those events without P60 detections, one
(GRB\,050607: \citealt{GCN.3527}) was detected in the optical below
our sensitivity limits, while three were detected in the NIR
(GRB\,050915: \citealt{GCN.3984}; GRB\,060923A: \citealt{GCN.5587};
GRB\,061222A: \citealt{GCN.5975}).  Only three events ($10\%$) in the entire
sample registered no detections in the optical or NIR bandpass: GRBs\,050412,
060805, and 070521.

$59\%$ of the events in the sample (17 of 29) have a redshift measured
from optical spectroscopy, roughly a factor of two larger than the \Swift\
population as a whole.  These range from $z = 0.6535$ (GRB\,050416A;
\citealt{snc+07}) to $z = 4.9$ (GRB\,060510B; \citealt{psc+07}).  Together 
with our measured median redshift of $\langle z \rangle \approx 2$, the events 
in our sample are relatively representative of \Swift\ afterglows ($\langle
z \rangle \approx 2.0$; \citealt{bkf+05,jlf+06}).  We wish to reiterate here
again that P60 immediately responds to all \Swift\ events visible at Palomar
Observatory (weather permitting), ruling out any large selection bias.  While 
small number statistics may account for some of our observed deviations from 
previous studies, our large detection efficiency merits a more thorough 
discussion in \S~\ref{sec:deteff:disc}.  

\subsection{X-Ray and Optical Flares}
\label{sec:xrayflares}
A large fraction ($\approx 33\%$) of \Swift\ X-ray light curves exhibit
dramatic short-lived flares superposed on their power-law decay \citep{fmr+07}.
The temporal and spectral structure of these flares indicate they cannot
come from the external shock powering the afterglow emission; instead they
are widely attributed to late-time activity of the central engine
\citep{zfd+06}.  Likewise, a re-brightening at late times in the optical 
bandpass has now been seen in several \Swift\ afterglows 
\citep{wvw+06,sdp+07}.  Investigating the relationship between these two
bandpasses should help shed light on the emission mechanisms responsible for 
these deviations from standard afterglow theory.

\begin{figure}[p]
  \epsscale{0.8}
  \centerline{\plotone{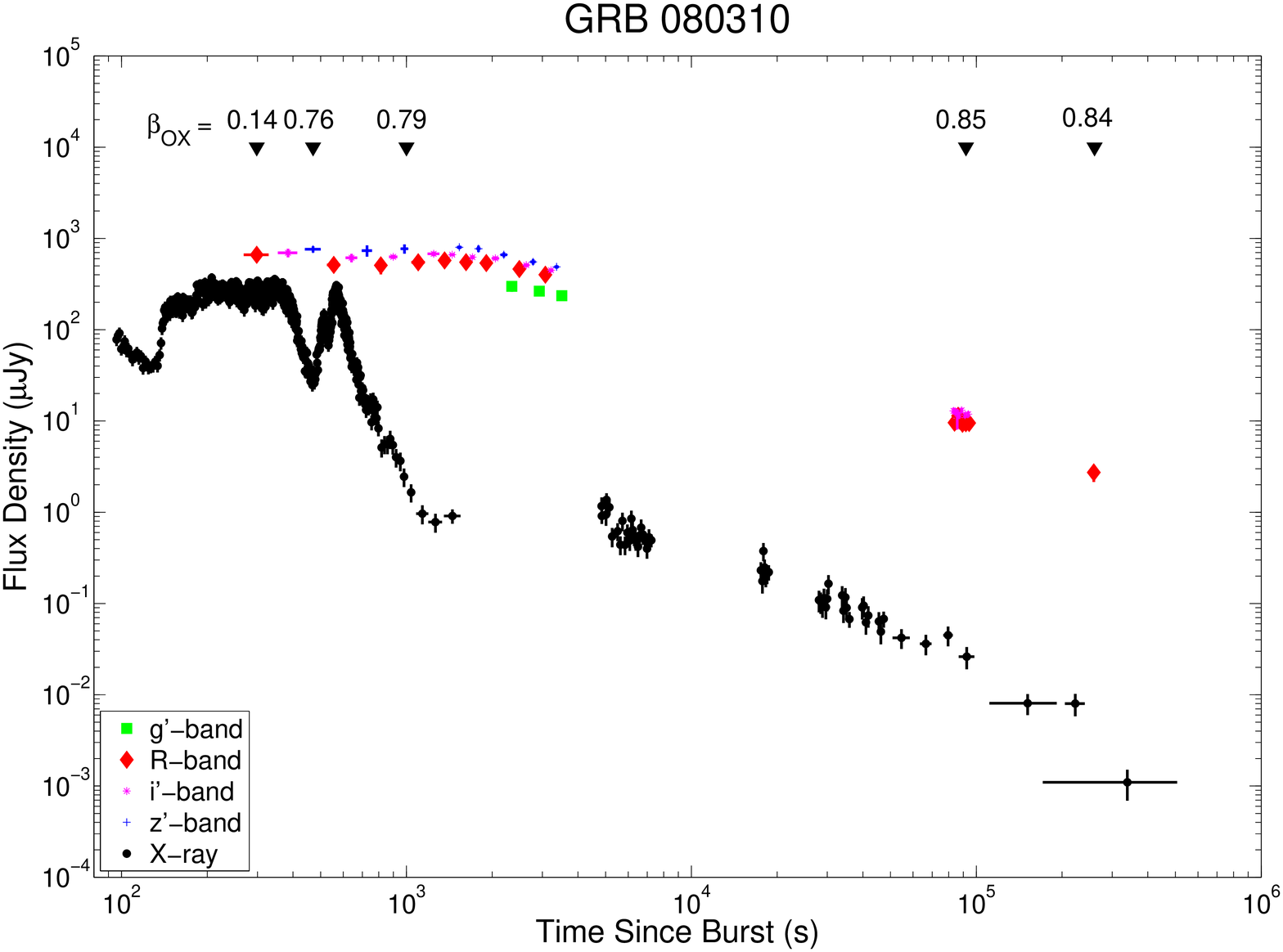}}
  \centerline{\plotone{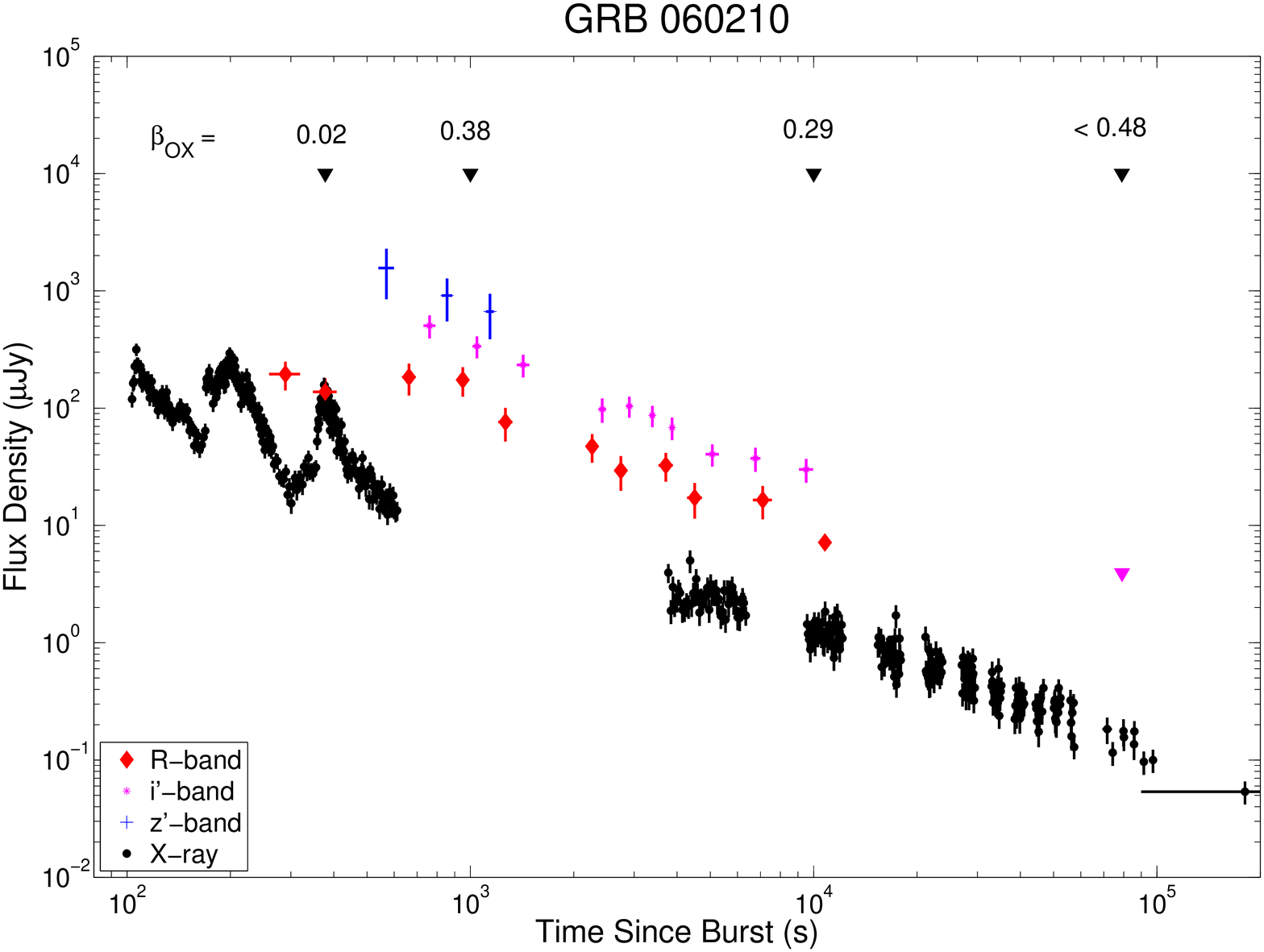}}
  \caption[X-ray and optical light curves of GRB\,080310 and GRB\,060210]
  {X-ray and optical light curves of GRB\,080310 ({\it top}) and GRB\,060210 
  ({\it bottom}).  For both events the optical light curve at early times 
  ($t \lesssim 10^{3}$\,s) is not correlated with the dramatic X-ray flares.  
  Measurement of the optical to X-ray spectral index ($\beta_{OX}$; 
  \S~\ref{sec:dark:obs}) is therefore a strong function of time.  Measuring 
  $\beta_{OX}$ during an X-ray flare may lead to erroneous classification of 
  some bursts as ``dark'' ($\beta_{OX} < 0.5$).  Both events, however, show 
  relatively constant $\beta_{OX}$ values for $t \gtrsim 10^{3}$\,s.  
  GRB\,060210, for example, is clearly a dark burst, even at late times.}
\label{fig:betaOX}
\end{figure}

\begin{figure}[p]
  \epsscale{0.8}
  \centerline{\plotone{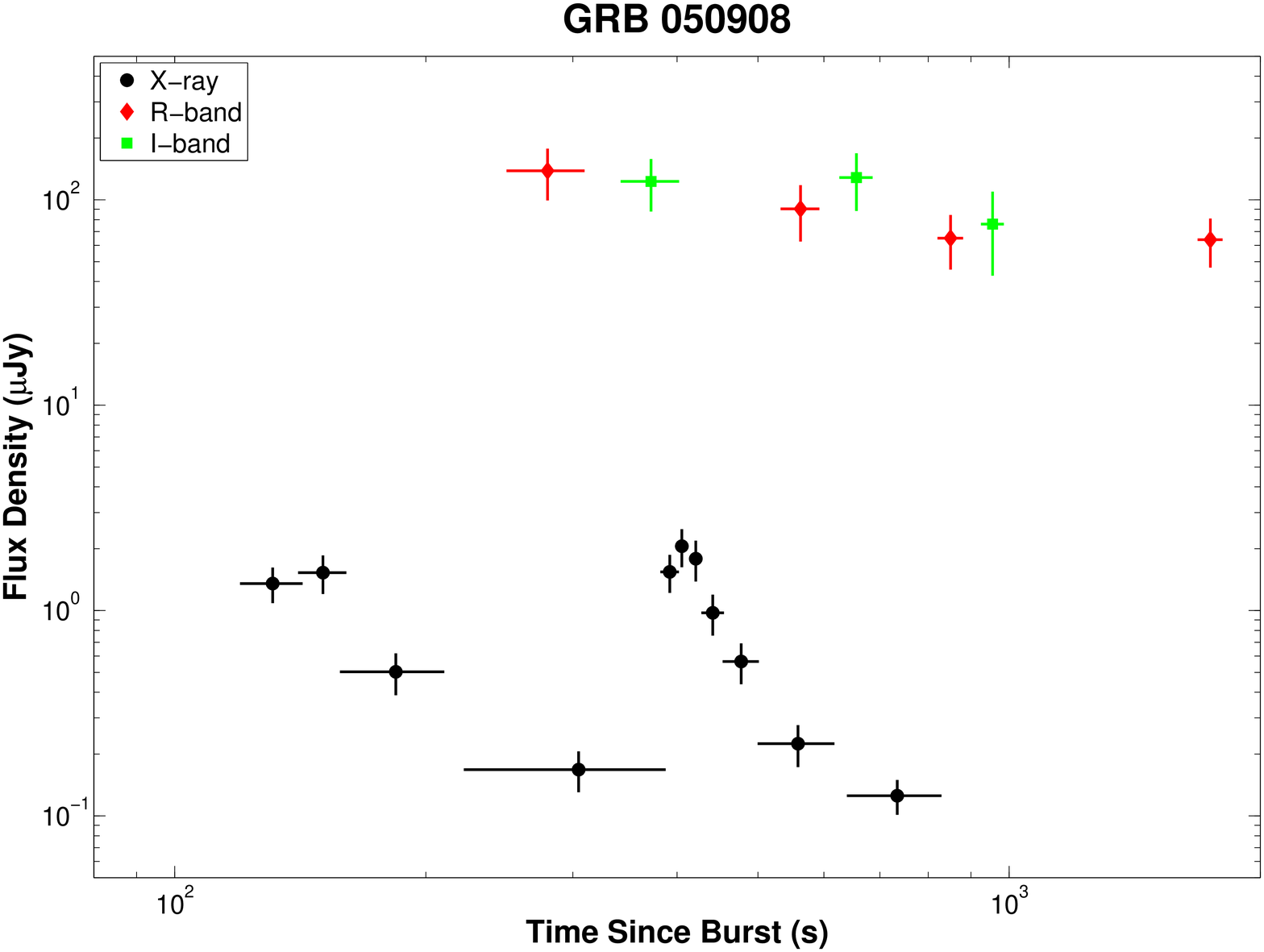}}
  \centerline{\plotone{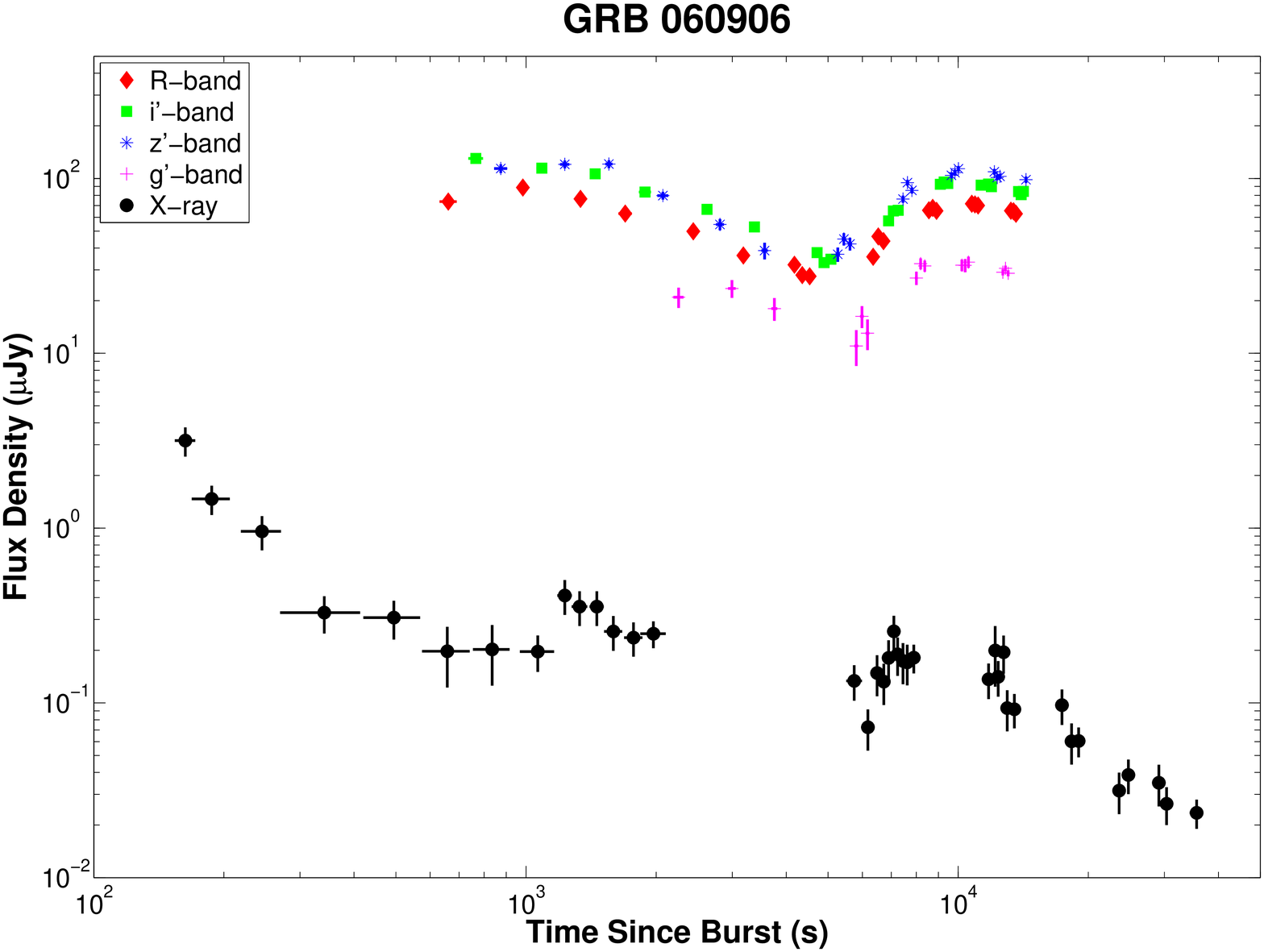}}
  \caption[X-ray and optical flares in \Swift\ afterglows]
  {X-ray and optical flares in \Swift\ afterglows.  \textit{Top:} X-ray and 
  optical light curves of GRB\,050908.  The X-ray light curve shows a dramatic 
  flare ($\Delta f / f \approx 50$ at $t \approx 400$\,s) at early times.  No 
  corresponding variability is seen in the optical.  \textit{Bottom:} X-ray 
  and optical light curves of GRB\,060906.  In this case, the re-brightening 
  occurs in the optical while the X-ray decay is relatively flat.  Both events 
  require additional emission mechanisms beyond the forward shock synchrotron 
  model.}
\label{fig:xrayflares}
\end{figure}

Our early afterglow sample includes four events with contemporaneous optical
observations of X-ray flares: GRBs\,050820A, 050908, 060210, and 080310
(Figs.~\ref{fig:betaOX} and \ref{fig:xrayflares}).  The relationship 
between the X-ray and optical emission from GRB\,050820A is discussed 
extensively in \citet{ckh+06} and \citet{vww+06}.  While the optical emission 
clearly jumps in concert with the bright X-ray flare at $t \approx 230$\,s, 
the dominant contribution to the optical emission at later times appears to 
come from the forward shock.  In the other three events, the optical emission 
is completely de-coupled from any flaring in the X-rays.

GRB\,060906 is unique in our sample, as we observe a re-brightening by a 
factor of $\approx$ 3 at $t \approx 10^{4}$\,s in the optical.  The 
X-ray decay, on the other hand, appears relatively flat during this stage.
One possibility to explain the
optical flare is an increase in the circumburst density; such a change in the
surrounding medium should have no effect on any emission above the 
synchrotron cooling frequency, $\nu_{c}$, where the X-ray bandpass is likely
to fall.  However, a recent study by \citet{ng07} has shown that even
sharp density changes do not lead to dramatic variability in afterglow
light curves; instead any changes in the afterglow evolution occurs smoothly
over several orders of magnitude in time.  We leave a more thorough discussion
of the afterglow of GRB\,060906 to Rana et al.~(2008, in preparation).

\subsection{Brightness and Luminosity Distribution}
\label{sec:luminosity:obs}
\begin{figure}[t!]
  \centerline{\plotone{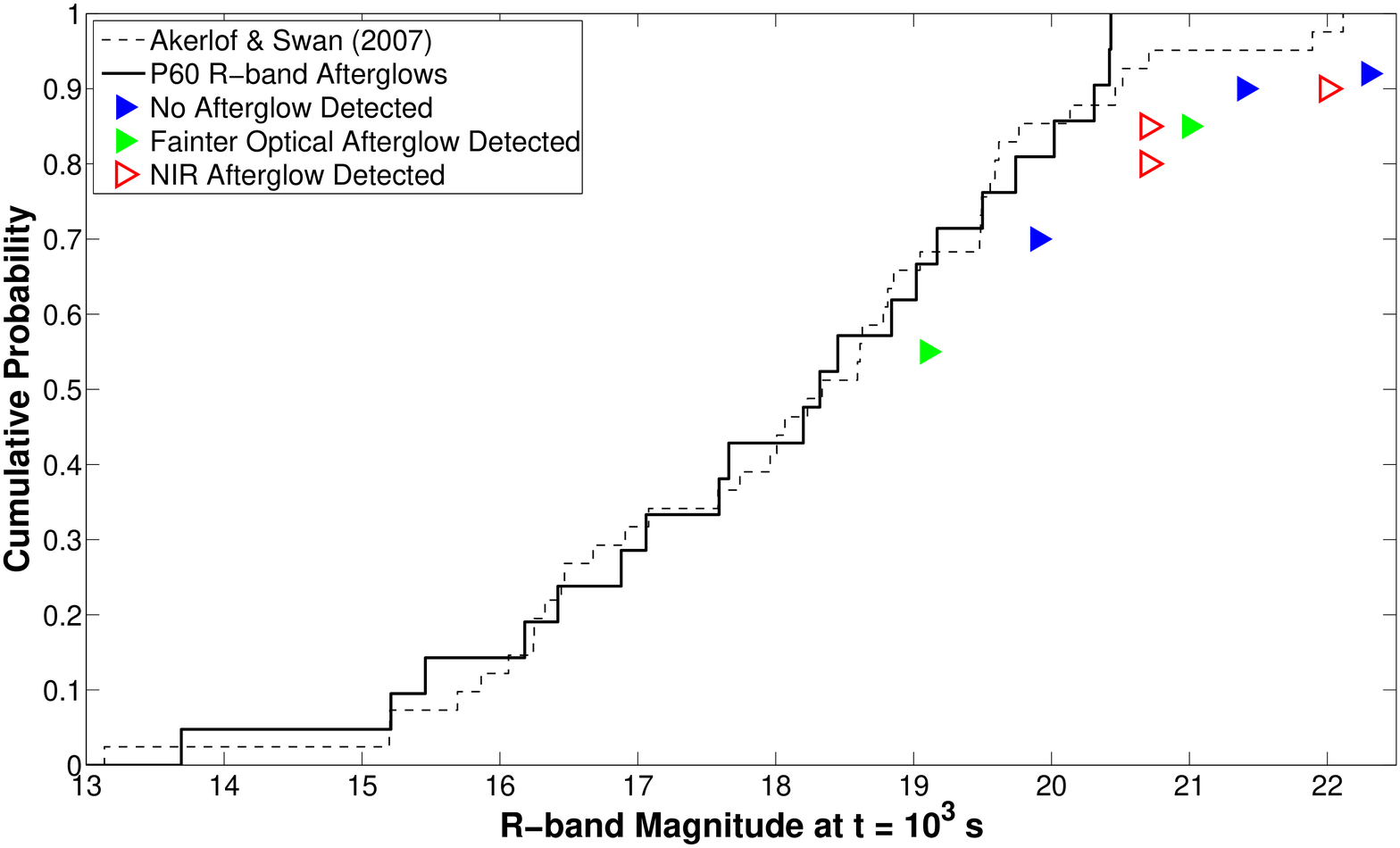}}
  \caption[P60-\Swift\ optical afterglow brightness distribution]
  {P60-\Swift\ optical afterglow brightness distribution.  We plot here the
  observed optical brightness distribution of all events in our early 
  afterglow sample at a common reference time of $t = 10^{3}$\,s (solid line).
  The dashed line indicates a similar archival analysis performed by 
  \citet{as07}.  Minor deviations can be seen at the very faint end 
  (\Rc $\gtrsim 21$\,mag), likely indicative of the P60 sensitivity limit.
  Arrows indicate P60 upper limits for GRBs with optical afterglows from other 
  facilities (green), GRBs with only NIR afterglows (red), and GRBs with no 
  detected optical or NIR afterglows (blue).}
\label{fig:agdistro}
\end{figure}

We have interpolated (where possible) or extrapolated the extinction-corrected
\Rc-band flux to a common time of $t = 10^{3}$\,s in the observer frame for 
21 P60-detected afterglows in our sample\footnote{GRB\,080320 was only detected
in the \ip\ and \zp\ filters and is therefore included in 
Figure~\ref{fig:agdistro} as a limit.}.  A plot of the resulting 
cumulative distribution is shown in Figure~\ref{fig:agdistro}.  For those 
events without detections, we take the deepest upper limit obtained before 
this fiducial time, and plot this limit as an arrow in 
Figure~\ref{fig:agdistro}.  For comparison, we also show the analogous result 
obtained by \citet{as07} in a literature-based study of the first 43 \Swift\ 
optical afterglows from 2005--2006.

It is clear from the large degree of overlap in the two distributions 
in Figure~\ref{fig:agdistro} that our sample, though slightly smaller in 
size, is consistent with the findings of \citet{as07} and therefore likely
representative of the entire \Swift\ optical afterglow population.  
We find a slight degree of variation at the faint end ($\Rc \gtrsim 
21.5$\,mag), which likely indicates we are missing a small fraction ($< 10\%$) 
of the faintest afterglows.  However, given that $\sim 70\%$ of \Swift\ events 
seem to have $\Rc < 22$\,mag at this fiducial time \citep{as07}, the P60 
sensitivity is well matched to detect the majority of events.

For those events for which we do not detect an optical afterglow with
P60, it is clear from Figures~\ref{fig:early:all} and \ref{fig:agdistro}
that only one event can be attributed to a lack of sensitivity (GRB\,050607,
which was located only 3\arcsec\ from a $R \approx 16$\,mag star).  
The remaining 6 events would all have been easily detected if as bright as
a typical afterglow in our sample.

\begin{figure}[t!]
  \centerline{\plotone{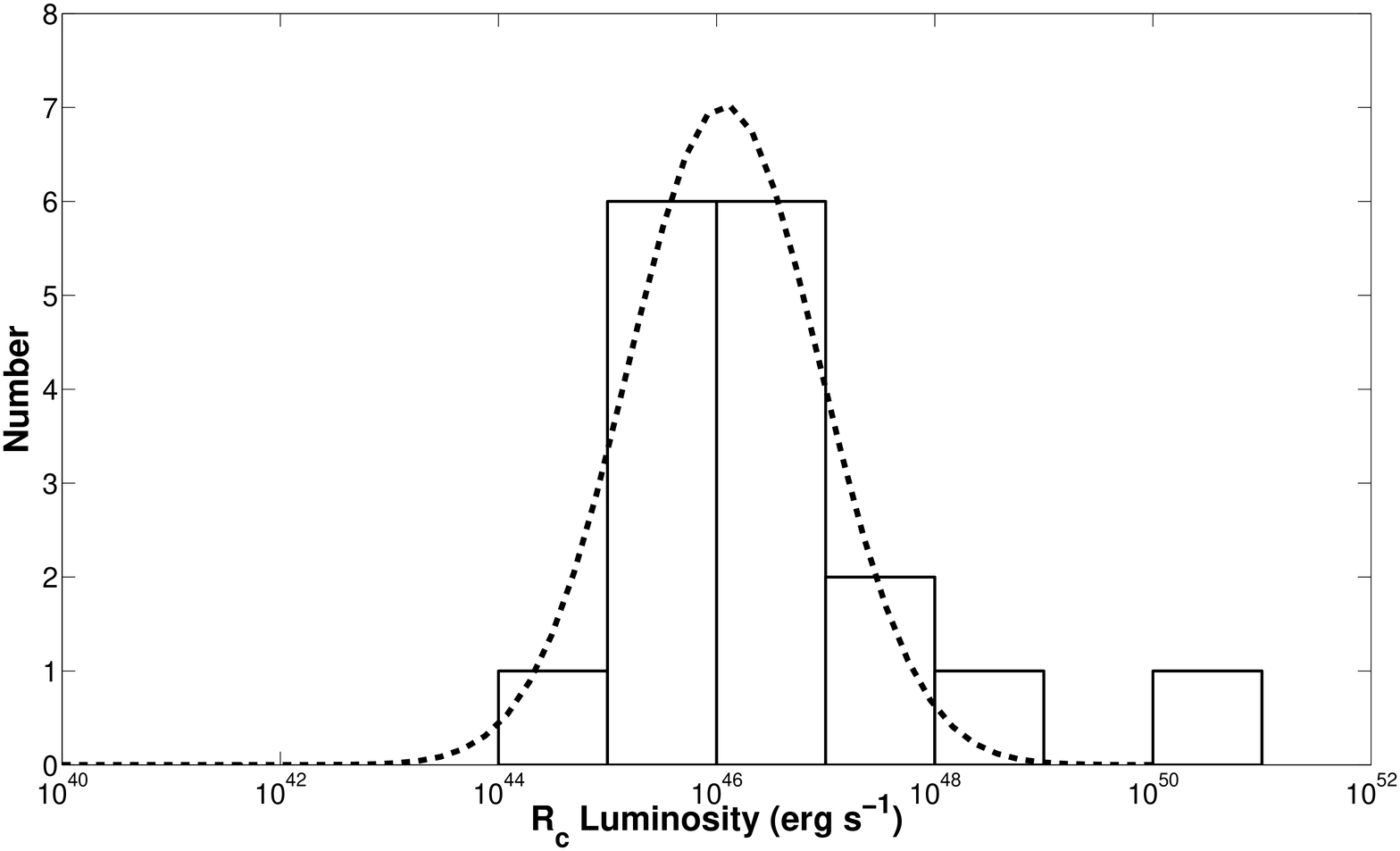}}
  \caption[P60-\Swift\ optical afterglow luminosity distribution]
  {P60-\Swift\ optical afterglow luminosity distribution.  We have measured
  the rest-frame optical \Rc-band luminosity at a common (rest frame) time
  of $t = 10^{3}$\,s for all events in our early sample with a spectroscopic
  redshift.  We find a good fit to a single log-normal distribution with mean
  $\log (L [\mathrm{erg s}^{-1}]) = 46.68$ and standard deviation 
  $\sigma = 1.04$ dex.  The sole outlier (GRB\,060210) falls on the 
  over-luminous end because of its extreme k-correction 
  (\S~\ref{sec:luminosity:obs}).}
\label{fig:agluminosity}
\end{figure}

For the 17 GRBs with redshifts, it is also possible to compare 
optical light curves in the GRB rest frame.  We therefore compute the 
afterglow luminosity at a fiducial time of $10^{3}$\,s in the rest frame of 
the GRB, applying a k-correction to convert our observed bandpass to the rest
frame $\Rc$-band, as described in \citet{hbb+02}.  The resulting 
histogram is shown in Figure~\ref{fig:agluminosity}.  At this
time, we find a median value for the afterglow luminosity to be
$\langle \log(L [\mathrm{erg s}^{-1}]) \rangle = 46.39$ with a standard 
deviation of 1.4 dex.  Also shown in Figure~\ref{fig:agluminosity} is the 
best-fit single Gaussian distribution.

Several authors \citep{lz06,ngg+06,kkz06} have argued in favor of a bimodal
distribution of intrinsic afterglow luminosity, with a class of nearby,
sub-luminous events.  Much like \citet{mmk+08}, we find no need for a 
bimodal distribution.  While a single event (GRB\,060210) is a significant
outlier on the over-luminous end, we note this event is at relatively high
redshift ($z = 3.91$; \citealt{GCN.4729}) and has an extremely steep
spectral index ($\beta = 7.2 \pm 0.7$).  The resulting k-correction is 
therefore extremely large (and relatively uncertain).  This seems a more likely
explanation than such an extremely luminous burst.
  
\subsection{Dark Bursts}
\label{sec:dark:obs}  
We adopt here the definition of a ``dark'' GRB as one where the optical 
(\Rc-band) to X-ray spectral index satisfies $\beta_{OX} < 0.5$ \citep{jhf+04}. 
Unlike definitions based solely on optical brightness, the $\beta_{OX}$ method
is physically motivated: an afterglow qualifies as dark when the
ratio of optical to X-ray flux is incompatible with standard synchrotron
afterglow theory.  By utilizing both the optical and X-ray afterglows, we can
easily distinguish between intrinsically sub-luminous afterglows (i.e., those
events that are faint in all bandpasses) and those afterglows that indicate
an additional process is selectively suppressing the optical flux (or, 
alternatively, increasing the X-ray emission).

\begin{figure}[t!]
  \centerline{\plotone{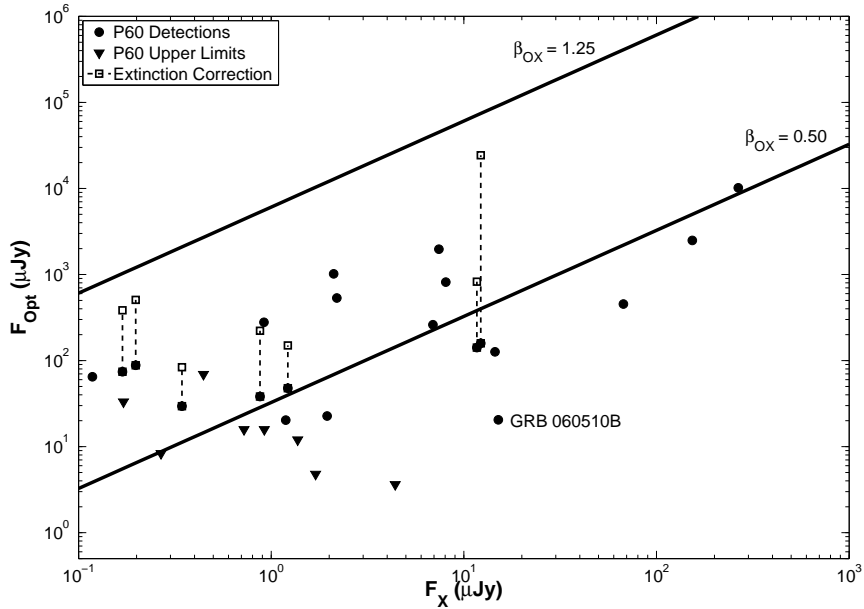}}
  \caption[Optical / X-ray spectral energy distribution of \Swift\ GRBs]
  {Optical / X-ray spectral energy distribution of \Swift\ GRBs.  We plot the
  X-ray and optical flux (or upper limits) at a common reference time of 
  $t = 10^{3}$ for all events in our P60-\Swift\ sample.  In
  standard afterglow theory, the optical to X-ray spectral index, $\beta_{OX}$
  should fall between $0.5 < \beta_{OX} < 1.25$ (solid black lines).  In our
  sample, nearly $50\%$ of afterglows quality as ``dark'' bursts ($\beta_{OX} <
  0.5$).  Correcting for extinction in the GRB host galaxy (\textit{open
  squares}) brings several events in line with the predictions of synchrotron
  radiation.}
\label{fig:darkbursts}
\end{figure}

In Figure~\ref{fig:darkbursts} we compare the X-ray and \Rc-band flux 
densities extrapolated to a common time of $t = 10^{3}$\,s for all 29 afterglows
in our sample.  The allowed region in the standard afterglow model, 
$0.50 \lesssim \beta_{OX} \lesssim 1.25$, is marked with solid lines.  Like
\citet{mmk+08}, we find that nearly $50\%$ of events qualify as dark under
this definition.  It is clear therefore that the faintness of the \Swift\
optical afterglows cannot be attributed solely to distance, as this would
not directly affect the measured flux ratio.  This result stands 
in stark contrast with the study of pre-\Swift\ events by 
\citet{jhf+04}, which found a dark burst incidence of only $10\%$.

The most important difference between our study and that of \citet{jhf+04} is
the time at which we evaluate $\beta_{OX}$ ($t = 11$\,hr for \citealt{jhf+04}).
In Figure~\ref{fig:betaOX} we demonstrate the importance of the reference
time when calculating $\beta_{OX}$.  Many \Swift\ afterglows exhibit bright
X-ray flares at early times \citep{brf+05}, as well as a plateau
decay phase indicative of continued energy injection into the forward shock
\citep{nkg+06,zfd+06}.  This late-time activity could artificially inflate
the X-ray flux at early times, leading to spuriously low $\beta_{OX}$
measurements (see GRB\,080310, Fig.~\ref{fig:betaOX}).

While our optical coverage at $t = 11$\,hr is relatively sparse, we find 
little evidence for evolution of $\beta_{OX}$ between these two epochs.
Delayed engine activity may explain the low values of $\beta_{OX}$ measured
for a few events (e.g., GRB\,050820A; \citealt{ckh+06}).  But we find that 
extrapolating the light curves out to both $t = 10^{4}$\,s and $t = 11$\,hr
does not change the dark burst fraction by more than 10$\%$.  This
echoes the result found by \citet{mmk+08}.

Another possibility to explain dark optical afterglows is a high-redshift
($z \gtrsim 5$) origin.  In this case, the observed \Rc\ bandpass falls below
the Ly-$\alpha$ cut-off in the GRB rest frame, leading to a significant
suppression of optical flux due to absorption in the intergalactic medium.
This is the case, for example, for GRB\,060510B ($\beta_{OX} = 0.04$), which
lies at $z = 4.9$ (Fig.~\ref{fig:darkbursts}; \citealt{psc+07}).  

Much like the delayed engine activity hypothesis, a high-redshift origin
can only account for a fraction of the observed dark bursts in our sample. 
Theoretical models, assuming GRBs trace the cosmic star formation rate,
predict a high-redshift ($z \gtrsim 7$) fraction of $\approx 10\%$
\citep{bl06}.  Five events with $\beta_{OX} < 0.5$ have measured 
spectroscopic redshifts, firmly establishing the Ly-$\alpha$ cut-off
below the observed \Rc\ filter (e.g., GRB\,060210: $\beta_{OX} = 0.37$, 
$z = 3.91$; \citealt{GCN.4729}).  And we can place upper limits on the
redshifts of some events that do not have optical afterglows based on the 
inference of absorption in excess of the Galactic value in X-ray afterglow 
spectra (e.g., GRB\,070521A: $z < 2.4$; \citealt{gnv+07}).

Finally, we consider the possibility of extinction native to GRB host
galaxies.  Because long-duration GRBs have massive star progenitors, it is
natural to expect them to explode in dusty, highly extinguished environments.
However, broadband studies of some of the best sampled afterglows in the
both pre-\Swift\ and \Swift\ eras indicate only a modest amount of host
reddening ($\langle A_{V} \rangle \approx 0.2$\,mag; \citealt{kkz06,kkz+07}).

In contrast, we find evidence for significant host absorption in several of 
the afterglows in our sample.  Using our multi-color P60 observations, we 
provide best-fit optical power-law spectral indices for all events with 
sufficient filter coverage in Table~\ref{tab:early}.  Of the 7 dark bursts 
with measured values of $\beta_{O}$, 6 spectral indices are too steep to be 
explained by the standard afterglow formulation (i.e., $\beta_{O} > 1.5$).  

To further quantify this effect, we have refitted our optical data, but in this
case fixing the optical spectral index to $\beta_{O} = 0.6$ (the average value
for bright \Swift\ events; \citealt{kkz+07}).  We then incorporated the effects
of dust by adding the host galaxy reddening [$A_{V}$(host)] as a free 
parameter to the fit.  In general, our data were not sufficient to distinguish
between competing extinction laws (i.e., Milky Way, LMC, and SMC;
\citealt{p92}).  We therefore assumed an SMC-like extinction curve, as this
model has proved successful for most GRB afterglows.  The results are shown in
Table~\ref{tab:early}.  The extinction-corrected fluxes are also plotted
in Figure~\ref{fig:darkbursts}.  In all cases where we were able to measure
the host extinction, this correction has moved the afterglow from near or
below the dark burst threshold back into the realm of synchrotron theory.

It is clear that the afterglows in our sample are significantly more reddened 
than the brightest afterglows in both the \Swift\ and pre-\Swift\ eras.
Furthermore, even our host absorption measurements are quite biased; we could
not measure $A_{V}$(host) for those events without P60 afterglows, which are
likely to be the most extinguished events in our sample.  Even some afterglows
that do not qualify as dark, such as GRB\,070208 and GRB\,070419A, exhibit
strong evidence for significant amounts of host galaxy dust [$A_{V}$(host)
$\approx 1$].  Though our sample size is still quite small, host galaxy
extinction appears to be the primary explanation for dark bursts in the 
\Swift\ era.
   
\section{Discussion and Conclusions}
\label{sec:discussion}


\subsection{Anomalous P60 Detection Efficiency}
\label{sec:deteff:disc}
We have demonstrated in \S~\ref{sec:deteff:obs} that P60 was able to detect
optical afterglow emission from a large fraction ($\sim 80\%$) of events for
which observations began within an hour of the burst trigger.  While the 
1.5\,m aperture is relatively large for a robotic facility, it would be
nonetheless informative to understand systematic effects that affect our
afterglow recovery rate.  The ultimate goal, of course, is to better inform 
future GRB follow-up campaigns.

The first lesson from this campaign is the importance of
observing in redder filters.  We have shown in 
\S~\ref{sec:dark:obs} that typical \Swift\ events suffer from a 
non-negligible amount of host galaxy extinction (Tab.~\ref{tab:early}).
Coupled with the additional effect of Ly-$\alpha$ absorption in the IGM from
a median redshift of $\langle z \rangle \approx 2$, it is clear that a large 
fraction of the low UVOT detection efficiency is caused by its 
blue observing bandpass.  The P60 automated follow-up sequence, consisting of
alternating exposures in the \Rc, \ip, and \zp\ filters, while initially
designed for identification of candidate high-$z$ events, is actually
well-suited to maximize afterglow detection rates.  

The large fraction of P60-detected bursts with spectroscopic redshifts, on the
other hand, is almost certainly an artifact of the unequal longitudinal
distribution of large optical telescopes.  With the exception of the South
African Large Telescope (SALT), all optical telescopes with apertures larger
than 8\,m fall within six time zones (UT-4 to UT-10).  It is not entirely
surprising then, that so many promptly discovered P60 optical 
afterglows have spectroscopic redshifts from immediate follow-up with the
largest optical facilities.

While building the largest optical facilities is often prohibitively
expensive for all but the largest collaborations, 1\,m class facilities are much
more feasible, both in terms of cost and construction time scale.  We wish here 
to echo the thoughts of many previous GRB observers (e.g., \citealt{as07}) that
future automated facilities be built at longitudes (and latitudes) not covered
by current facilities.  NIR coverage is particularly crucial to detect the 
most extinguished events and provide tighter constraints on the afterglow
SED and hence host galaxy extinction.

A longitudinally spaced ring of 1\,m class facilities, as for example envisioned
by the Las Cumbres Observatory Global Telescope\footnote{See http://lcogt.net.}
is well positioned in the future to recover the vast majority of GRB optical
afterglows, assuming the follow-up is done in the reddest filters possible.
Such coverage will be particularly important as we transition into the 
\textit{Fermi} era, with its significantly decreased rate of precise GRB 
localizations.

\subsection{Re-visiting Dark Bursts}
\label{sec:darkbursts:disc}
We now turn our attention to the issue of dark bursts in the \Swift\ era.
In \S~\ref{sec:dark:obs}, we demonstrated that a large fraction
($\approx 50\%$) of \Swift\ afterglows showed suppressed emission in the
optical bandpass (relative to the X-ray), that was due in large part to
extinction in the host galaxy.  Given the natural expectation that GRBs, 
since they are associated with massive stars, should form in relatively
dusty environments, we wish to understand why our study of \Swift\
events yields such a dramatically different dark burst fraction than previous
work on pre-\Swift\ GRBs \citep{jhf+04}.

We believe selection effects are one large cause of this discrepancy.
It is clear that previous studies of GRB host galaxy extinction, by selecting 
the brightest and best-sampled events, provide a strongly biased view.  Many,
if not most, GRB hosts, appear to suffer from a significant amount of dust
extinction ($A_{V} \gtrsim 0.5$).  Even the study of \citet{jhf+04}, though
it included \textit{all} pre-\Swift\ GRBs with an X-ray afterglow, could 
be biased towards unextinguished events as well.  Before \Swift, 
target-of-opportunity X-ray observations often required the accurate 
localization provided by an optical (or radio) afterglow.  Thus those events 
with the brightest optical afterglows (assumed to have on average smaller 
extinction) were more likely to be observed in the X-ray, biasing the 
optical-to-X-ray spectral index to larger values of $\beta_{OX}$.         

Another, more subtle, effect, may also cause \Swift\ afterglows to appear 
darker than pre-\Swift\ afterglows, independent of host galaxy extinction.
Because \Swift\ is a more sensitive instrument, it detects GRBs at a higher
average redshift than any previous mission.  Consider a host frame extinction 
of $A_{V} = 0.1$\,mag.  At $z = 1$, typical for pre-\Swift\ events, the observed
\Rc\ filter corresponds to roughly to rest-frame $U$-band, and so an
extinction of 0.17\,mag (assuming a Milky Way-like extinction curve).  On
the other hand, at $z = 3$, the observed \Rc-band corresponds to a rest frame
wavelength of $\lambda = 1647$\,\AA.  So at high redshift, the same amount of 
dust will produce nearly twice as much extinction in the observed bandpass.
Solely because of redshifts effects, similar environments will produce 
different observed spectral slopes.  This effect is exacerbated by the nature
of dust grains in most GRB host galaxies, as the SMC extinction curve does not
show the pronounced 2175\,\AA\ bump seen from the Milky Way \citep{p92}.

If GRBs do trace the cosmic star formation rate, our results suggest a 
significant fraction of star formation occurs in highly obscured environments.
\citet{kkz06} found a weak correlation between host reddening and sub-mm
flux, and we believe a sensitive mid-infrared or sub-mm survey of GRB host 
galaxies would be an important confirmation of our results.  However, instead
of focusing on the brightest, best studied afterglows, as has often been
done in the past (e.g., \citealt{tbb+04,mhc+08}), we instead suggest a survey 
of the host galaxies of the optically darkest GRB afterglows to see if these 
events really do exhibit signs of obscured star formation.

\acknowledgments
P60 operations are funded in part by NASA through the \textit{Swift} Guest
Investigator Program (Grant Number NNG06GH61G).  S.~B.~C.~is supported by 
a NASA Graduate Student Research Program fellowship.  M.~M.~K.~would like
to acknowledge the Moore Foundation for the Hale Fellowship supporting her
graduate studies.  A.~G.~acknowledges support by the Benoziyo Center for 
Astrophysics and the William Z.~and Eda Bess Novick New Scientists Fund at the 
Weizmann Institute.  This work made use of data supplied by the UK 
\textit{Swift} Science Data Centre at the University of Leicester.  This 
research has made use of the USNOFS Image and Catalogue Archive operated by 
the United States Naval Observatory, Flagstaff Station.

{\it Facilities:} \facility{PO:1.5m ()}, \facility{Swift ()}


\clearpage
\input{tab1.tex}

\clearpage
\input{tab2.tex}

\clearpage
\input{tab3.tex}

\end{document}